\documentclass[twocolumn,aps,prl,preprintnumbers]{revtex4}
\usepackage{bm}
\usepackage{bbm}
\usepackage[latin9]{inputenc}
\usepackage{amsmath}
\usepackage{amssymb}
\usepackage{graphicx}
\usepackage{mathrsfs}
\usepackage{amsfonts}
\usepackage{amsthm}
\usepackage{color}
\usepackage{ulem}
\usepackage{txfonts}
\usepackage[colorlinks=true,citecolor=blue,linkcolor=blue,urlcolor=blue,anchorcolor=blue]{hyperref}%
\usepackage{pifont}
\hypersetup{colorlinks=true,citecolor=blue,linkcolor=blue,urlcolor=blue}
\providecommand{\U}[1]{\protect\rule{.1in}{.1in}}
\makeatletter
\@ifundefined{textcolor}{}
{
	\definecolor{BLACK}{gray}{0}
	\definecolor{WHITE}{gray}{1}
	\definecolor{RED}{rgb}{1,0,0}
	\definecolor{GREEN}{rgb}{0,1,0}
	\definecolor{BLUE}{rgb}{0,0,1}
	\definecolor{CYAN}{cmyk}{1,0,0,0}
	\definecolor{MAGENTA}{cmyk}{0,1,0,0}
	\definecolor{YELLOW}{cmyk}{0,0,1,0}
}

\makeatother
\begin{document}

\title{Alignment-Dependent Gapless Chiral Split Magnons in Altermagnetic Domain Walls}

\author{Zhaozhuo Zeng}
\author{Zhejunyu Jin}
\author{Peng Yan}
\email[Contact author: ]{yan@uestc.edu.cn}
\affiliation{School of Physics and State Key Laboratory of Electronic Thin Films and Integrated Devices, University of Electronic Science and Technology of China, Chengdu 611731, China}

\begin{abstract}
Altermagnets, an emerging class of magnetic materials, exhibit exotic chiral split magnons that are of great interest for both fundamental physics and spintronic applications. However, detecting and manipulating these magnons is challenging due to their THz frequency response. Here, we report the discovery of gapless chiral split magnons confined within altermagnetic domain walls. Unlike in conventional ferromagnets or antiferromagnets, their spectrum is highly sensitive to the domain wall orientation relative to the crystal axis. These magnons inherit the chiral splitting of their bulk counterparts and are detectable in the microwave regime, offering a distinctive signature for identifying altermagnets. We further show that the interfacial Dzyaloshinskii-Moriya interaction drives hybridization of magnons with opposite chiralities, enabling unidirectional strong magnon-magnon coupling. Moreover, we demonstrate that spin-orbit torque can control the domain wall orientation, providing a practical means to manipulate these chiral magnons. Our findings open pathways for novel magnonic nanocircuitry based on altermagnetic domain walls.

\end{abstract}

\maketitle

\textit{Introduction---}Altermagnets represent a new phase of magnetism that have captivated researchers due to their extraordinary properties. These materials combine spin splitting in their electronic band structure with a staggered antiferromagnetic order in real space \cite{vsmejkal2022beyond,vsmejkal2022emerging,mazin2022altermagnetism,bai2024altermagnetism,song2025altermagnets}. Unlike traditional antiferromagnets, altermagnets exhibit unique electrical phenomena, such as the anomalous Hall effect \cite{vsmejkal2020crystal,gonzalez2023spontaneous,reichlova2024observation}, anomalous Nernst effect \cite{badura2024observation,han2024observation}, giant tunnelling magnetoresistance \cite{vsmejkal2022giant,liu2024giant}, and non-relativistic spin polarization \cite{naka2019spin,bai2022observation,karube2022observation}. Their potential for ultrafast magnetization dynamics further positions them as candidates for next-generation spintronic technologies \cite{bai2024altermagnetism,baltz2018antiferromagnetic}. A defining characteristic of altermagnets is their anisotropic chiral split magnon bands, where magnons (the quanta of spin waves) of opposite chiralities exhibit direction-dependent frequency splitting \cite{vsmejkal2023chiral,cui2023efficient,liu2024chiral,morano2025absence,gomonay2024structure}. Recent studies suggest this chiral splitting could serve as a hallmark of altermagnetism \cite{krempasky2024altermagnetic,amin2024nanoscale,ding2024large,hariki2024xray}. Despite their promise, identifying altermagnets experimentally remains difficult. Efforts to detect spin splitting in electronic bands have yielded inconsistent results \cite{hiraishi2024nonmagnetic,liu2024absence,noh2025tunneling}, while the anomalous Hall effect, weakened by factors like self-doping and spin canting, proves unreliable as a standalone indicator. The chiral split magnon bands, though a promising signature, are typically in the THz frequency range, requiring advanced techniques like inelastic neutron scattering for detection \cite{liu2024chiral,morano2025absence}. This limitation complicates routine characterization and restricts broader exploration of altermagnetic behavior.

To overcome these obstacles, we aim to develop an accessible approach for detecting and manipulating chiral magnons in altermagnets. Specifically, we seek to bring their detection into the microwave frequency range, where widely available tools like all-electrical spin-wave spectroscopy or Brillouin light scattering \cite{yu2021magnetic} can be employed. Such a shift would simplify the identification of altermagnets and unlock opportunities to study their magnonic properties for practical applications \cite{jin2025strong}. The distinctive properties of altermagnets arise from their broken parity-time $\left(\mathcal{PT}\right)$ symmetry \cite{song2025altermagnets}. This leads to anisotropic exchange interactions that couple magnon chirality to the crystal axes, in contrast to ferromagnets (only one type of chirality) and collinear antiferromagnets (degenerate magnon modes without external magnetic fields). In domain walls (DWs), the translational symmetry in the direction perpendicular to the wall is broken, but it is preserved along the wall. This broken symmetry gives rise to a Goldstone mode \cite{Goldstone2022}: a gapless excitation corresponding to rigid translation of the DW position. The anisotropy however enables orientation-dependent magnon behaviors, further modulated by interactions like Dzyaloshinskii-Moriya interaction (DMI), which are absent in conventional magnetic systems.

\begin{figure}[t]
	\includegraphics[width=1\columnwidth]{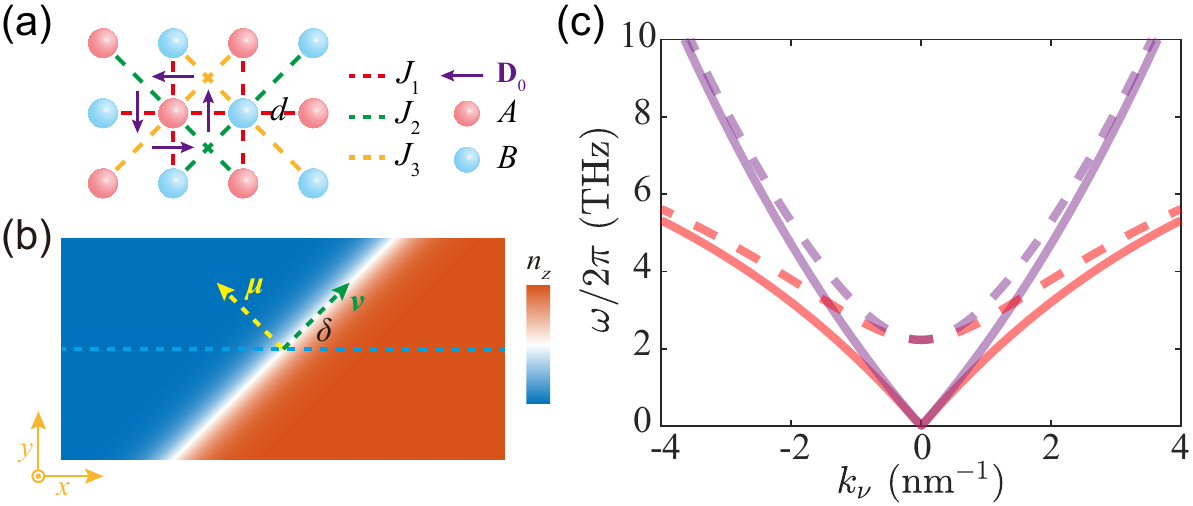}%
	\caption{\label{Fig:1}(a) The schematic of two-dimensional atomistic spin model. $J_1$, $J_2$, and $J_3$ represent the distinct Heisenberg exchange interactions. $ {\bf{D}}_0 $ is the DMI vector. $ A $ and $ B $ label two sublattices. $ d $ is the lattice constant. (b) The distribution of the $ z $-component of N\'{e}el vector $ n_z $ with an oblique AMDW. $ \bm{\mu} $ and $ \bm{\nu} $ are the normal and tangential directions of the wall, respectively. $ \delta $ is the oblique angle of domain wall. (c) The dispersions of magnons along the wall with $\delta = 45^\circ$ and $D_0=0$. Solid and dotted lines represent the bound [Eq.~\eqref{eq:8}] and bulk [Eq.~\eqref{eq:5}] magnons, respectively. Red and purple indicate the RH and LH chirality, respectively.}  
\end{figure}

In this Letter, we explore the behavior of chiral magnons confined within altermagnetic DWs (AMDWs) \cite{gomonay2024structure,kolezhuk2024current,zarzuela2025transport,kravchuk2025chiral}, which function as one-dimensional waveguides \cite{garcia2015narrow,zhang2018eavesdropping,yu2018polarization,henry2019unidirectional,park2019channeling,wang2020stacking,chen2020narrow,qiu2022tunable,liang2022nonreciprocal,li2024realizing}. Using a two-dimensional atomistic spin model [see Fig.~\ref{Fig:1}(a)], we focus on $d$-wave AMDWs and uncover several striking features: (i) Gapless bound magnons: unlike typical magnetic systems, these magnons energy gaps; (ii) Alignment-dependent dispersions: because of the anisotropic exchange interaction, the right-handed (RH) and left-handed (LH) chiral magnons display dispersions that vary with the AMDW's orientation relative to the crystal axis, adhering to [$C_2||C_{4z}$] symmetry. This behavior differs from ferromagnetic or antiferromagnetic domain walls \cite{wagner2016magnetic,albisetti2018nanoscale,sluka2019emission,chen2025observation}, where dispersions are either orientation-independent or lack chiral distinction without external fields; (iii) Unidirectional coupling: the DMI enables a unique, one-way interaction between RH and LH magnons, a feature exclusive to altermagnetic textures; (iv) Electrical control: the orientation of AMDWs can be adjusted using spin-orbit torque (SOT) and/or spin-transfer torque (STT), offering a practical method to tune these magnons. Our results provide a microwave-accessible signature for altermagnets, bypassing the challenges of THz detection and ambiguous electrical signals. By revealing the distinctive transport properties of chiral magnons in AMDWs, this work enhances our understanding of altermagnetism and lays the groundwork for innovative magnonic technologies exploiting their controllable, low-frequency magnon modes.

\textit{Model---}We start with a two-dimensional atomistic spin model of altermagnet \cite{cui2023efficient}, shown in Fig.~\ref{Fig:1}(a). The continuum Hamiltonian with dimensionless N\'{e}el vector $ {\bf{n}} = ({\bf{m}}_{A}-{\bf{m}}_{B})/2 $ and magnetization vector $ {\bf{m}} = ({\bf{m}}_{A}+{\bf{m}}_{B})/2 $ is given by (see Sec. I in Supplemental Material \cite{SM})
\begin{equation}
	\begin{aligned}
	\mathcal{H}=&\int\left\{A_{0}{\bf{m}}^{2}+A_{1}\left(\nabla{\bf{n}}\right)^{2}+A_{2}\left(\partial_{x}{\bf{m}}\cdot\partial_{y}{\bf{n}}+\partial_{x}{\bf{n}}\cdot\partial_{y}{\bf{m}}\right)\right.\\
	&\left.-Kn_{z}^{2}+D\left[\left({\bf{n}}\cdot\nabla\right)n_{z}-n_{z}\left(\nabla\cdot{\bf{n}}\right)\right]\right\} d{\bf{r}},
	\end{aligned} \label{eq:1}
\end{equation}
where $ A_0  = 4J_1S^2/d^3 $ and $ A_1  = (J_1-J_2-J_3)S^2/\left(2d\right) $ are the homogeneous and inhomogeneous exchange parameters, $ A_2 = (J_2-J_3)S^2/d $ is the altermagnetic exchange constant, $ K = K_0S^2/d^3 $ and $ D = D_0S^2/d^2 $ are the anisotropy and DMI coefficients, respectively, and $ d $ is the lattice constant. In the present model, $J_1 > 0$ and $J_{2,3} < 0$ label the nearest-neighboring antiferromagnetic and next-nearest-neighboring ferromagnetic coupling strength, respectively, $ K_0 > 0 $ is the easy-axis magnetic anisotropy, $ {D}_{0} $ is the interfacial DMI \cite{fert2013skyrmions,jin2024skyrmion,vakili2025spin}, and $ S $ is the magnitude of spin.

We then obtain the equation of motion for the N\'{e}el vector $ {\bf{n}} $
\begin{equation}
	\begin{aligned}
	&  \frac{A_{2}}{A_{0}}{\bf{n}}\times\left(\dot{{\bf{n}}}\times\partial_{xy}^{2}{\bf{n}}+\partial_{x}{\bf{n}}\times\partial_{y}\dot{{\bf{n}}}+\partial_{y}{\bf{n}}\times\partial_{x}\dot{{\bf{n}}}+2{\bf{n}}\times\partial_{xy}^{2}\dot{{\bf{n}}}\right)\\
	&+\frac{M_{s}}{\gamma A_{0}}{\bf{n}}\times\ddot{{\bf{n}}}-\gamma{\bf{n}}\times {\bf{f}}_n =0, 
	\end{aligned} \label{eq:2}
\end{equation}
where $ {\bf{f}}_n = \left\{ A_{1}\nabla^{2}{\bf{n}}+Kn_{z}{\bf{e}}_z+D\left[\left(\nabla\cdot{\bf{n}}\right){\bf{e}}_z-\nabla n_{z}\right]\right\} /M_{s}$, $ M_s = \mu_s/\left(2d^3\right) $ is saturation magnetization with $ \mu_s = g\mu_\mathrm{B}S $ being the atomic magnetic moment, $ g $ is the is the Land\'e factor, $ \mu_\mathrm{B} $ is the Bohr magneton, and $ \gamma $ is the gyromagnetic ratio. By substituting $ {\bf{n}} = \sin\theta\cos\varphi{\bf{e}}_x+\sin\theta\sin\varphi{\bf{e}}_y+\cos\theta{\bf{e}}_z $ into Eq.~\eqref{eq:2} with the polar angle $ \theta(x,y) $ and azimuthal angle $ \varphi(x,y) $, we obtain the governing equation for the static profile of altermagnets
\begin{equation}
	\begin{aligned}
	A_{1}\nabla^{2}\theta-K\sin\theta\cos\theta&=0,\\
	\sin\varphi\partial_{x}\theta-\cos\varphi\partial_{y}\theta&=0.
	\end{aligned} \label{eq:3}
\end{equation}
By considering the boundary conditions $ \theta\left|_{\mu=+\infty\left(-\infty\right)}\right. = \pi\left(0\right) $, we obtain the Walker solution $ \theta=2\arctan\left[\exp\left(\mu/w\right)\right]=2\arctan\left[\exp\left(-x\sin\delta/w+y\cos\delta/w\right)\right] $ of AMDWs, where $ \mu $ is the coordinate along the normal direction of DW, $ w=\sqrt{A_{1}/K} $ is the DW width, and $ \delta $ is the oblique angle of DW, as illustrated in Fig.~\ref{Fig:1}(b). The azimuthal angle $ \varphi=\delta+\pi/2 $. Interestingly, despite the anisotropic exchange interactions in altermagnets, the DW width is determined solely by $ A_1 $ and $ K $, and is independent of its orientations.

In the local frame, the N\'{e}el vector $ {\bf{n}} $ can be divided into the static background $ n_3 {\bf{e}}_3 $ and fluctuations $ n_1 {\bf{e}}_1 + n_2 {\bf{e}}_2 $, where $ {\bf{e}}_3 = {\bf{e}}_1 \times {\bf{e}}_2 $, $ \left|n_{1}\left(x,y,t\right)\right|,\left|n_{2}\left(x,y,t\right)\right|\ll1 $, and $ n_{3}\approx1 $. By employing a three-dimensional rotation $ \mathcal{R} = \exp(\varphi L_z)\exp(\theta L_y) $ \cite{kim2019tunable,jin2023nonlinear}, we can transform the $ z $ axis into the static background $ {\bf{e}}_3 $, i.e., $ \mathcal{R} {\bf{e}}_z = {\bf{e}}_3 = \sin\theta\cos\varphi{\bf{e}}_x+\sin\theta\sin\varphi{\bf{e}}_y+\cos\theta{\bf{e}}_z $. Here, $ L_y $ and $ L_z $ are generators of the rotations with respect to the $ y $ and $ z $ axes, respectively. We thus derive the coupled dynamic equations under the basis of magnon fluctuations $ \left(n_1,n_2\right)^\top $
\begin{equation}
	\left(\begin{array}{cc}
	\mathcal{T}-\mathcal{B}_{1} & \mathcal{B}_{2}-\mathcal{D}_{1}\\
	-\mathcal{B}_{2}+\mathcal{D}_{1} & \mathcal{T}-\mathcal{B}_{1}+\mathcal{D}_{2}
	\end{array}\right)\left(\begin{array}{c}
	n_{1}\\
	n_{2}
	\end{array}\right)=0, \label{eq:4}
\end{equation}
where $ \mathcal{T} = M_{s}/\left(\gamma A_{0}\right)\partial_{tt}^2 $, $ \mathcal{B}_{1} = \gamma A_{1}/M_{s}\left(\partial_{\mu\mu}^{2}+\partial_{\nu\nu}^{2}\right)-\gamma K \cos2\theta/{M_{s}} $, $ \mathcal{B}_{2} = A_{2}K/\left(2A_{0}A_{1}\right)\sin\left(2\delta\right)\sin^{2}\theta\partial_{t}+A_{2}\sin\left(2\delta\right)/{A_{0}}\left(\partial_{\mu\mu}^{2}-\partial_{\nu\nu}^{2}\right)\partial_{t}-2A_{2}\cos\left(2\delta\right)/{A_{0}}\partial_{\mu\nu}^{2}\partial_{t}$, $ \mathcal{D}_{1} = \left(\gamma D/{M_{s}}\right)\sin\theta\partial_{\nu} $, and $ \mathcal{D}_{2} = \gamma D \sin\theta/\left({M_{s}w}\right) $. For bulk magnons propagating along the tangential direction $ \bm{\nu} $ with the eigenfunction $ n_{1(2)} = n_{1(2)_0}\exp\left(i{\bf{k}}_{\nu}\cdot{\bm{\nu}}-i\omega t\right) $, we obtain
\begin{equation}
	\omega_{\pm}^{\mathrm{bulk}}=\frac{\sqrt{\left[A_{2}\sin\left(2\delta\right)k_{\nu}^{2}\right]^{2}+4A_{0}\left(A_{1}k_{\nu}^{2}+K\right)}\pm A_{2}\sin\left(2\delta\right)k_{\nu}^{2}}{2M_{s}/\gamma}, \label{eq:5}
\end{equation}
where the positive (negative) sign represents the RH (LH) magnon mode. As for the bound magnons in AMDWs, we consider a degenerate perturbation theory with the eigenfunction $ n_{1(2)} = n_{1(2)_0}\exp\left(i{\bf{k}}_{\nu}\cdot{\bm{\nu}}-i\omega t\right) \mathrm{sech}\left(\mu/{w}\right) $ \cite{li2024realizing}. Subsequently, we have (see Sec. II \cite{SM})
\begin{equation}
	\left(\begin{array}{cc}
	T\omega^2-B_{1}k_{\nu}^{2} & i\omega B_{2}k_{\nu}^{2}-iD_{1}k_{\nu}\\
	-i\omega B_{2}k_{\nu}^{2}+iD_{1}k_{\nu} & T\omega^2-B_{1}k_{\nu}^{2}+D_{2}
	\end{array}\right)\left(\begin{array}{c}
	n_{1}\\
	n_{2}
	\end{array}\right)=0, \label{eq:6}
\end{equation}
where $ {T} = -M_{s}/\left(\gamma A_{0}\right) $, $ B_{1} = -\gamma A_{1}/{M_{s}} $, $ B_2 = -A_{2}\sin\left(2\delta\right)/{A_{0}} $, $ D_1 = \gamma D\pi/\left(4M_{s}\right) $, and $ D_2 = \gamma D\pi/\left(4M_{s}w\right) $. The secular equation \eqref{eq:6} gives
\begin{equation}
	\left(T\omega^{2}-B_{1}k_{\nu}^{2}\right)\left(T\omega^{2}-B_{1}k_{\nu}^{2}+D_{2}\right)-\left(\omega B_{2}k_{\nu}^{2}-D_{1}k_{\nu}\right)^{2}=0, \label{eq:7}
\end{equation}
the solution of which gives rise to the dispersion of bounded magnons. To distinguish the magnon chirality, we define the parameter $ \varepsilon=-in_2/n_1=\left(T\omega^2-B_{1}k_{\nu}^{2}\right)/\left(\omega B_{2}k_{\nu}^{2}-D_{1}k_{\nu}\right) $ \cite{jiao2024universal}. It is found that $ \varepsilon > 0 $, $ \varepsilon < 0 $, and $ \varepsilon = 0 $ or $\infty$ correspond to RH, LH, and linear magnon polarizations, respectively.

To examine our analytical results, we perform atomistic spin dynamics simulations by solving the Landau-Lifshitz-Gilbert equation (see details in Sec. III \cite{SM}). The magnetic parameters of altermagnetic insulator $ \mathrm{Cr_2Te_2O} $ \cite{cui2023efficient} are adopted: $ J_1 = 3.90 $ meV, $ J_2 = -7.90 $ meV, $ J_3 = -1.21 $ meV, $ K_0 = 0.73 $ meV, $ \mu_s = 3.29\mu_{\mathrm{B}} $, $ S = 3/2 $, $ d = 0.291 $ nm, and Gilbert damping constant $ \alpha = 10^{-3} $, giving rise to an ultra-narrow DW width $ w=0.87 $ nm.

\textit{No DMI---}We first consider the case without the DMI, i.e., $ D_0 $ = 0. The solution of Eq.~\eqref{eq:7} is reduced to
\begin{equation}
	\omega_{\pm}^{\mathrm{bound}}=\frac{\sqrt{\left[A_{2}\sin\left(2\delta\right)k_{\nu}^{2}\right]^{2}+4A_{0}A_{1}k_{\nu}^{2}}\pm A_{2}\sin\left(2\delta\right)k_{\nu}^{2}}{2M_{s}/\gamma}. \label{eq:8}
\end{equation} 
From the above formula, we find that the bound magnons are gapless, i.e., $\omega_{\pm}^{\mathrm{bound}}(k_{\nu}=0)=0$, for all angles $\delta$. Numerical results in Fig.~\ref{Fig:1}(c) confirm this point for a specific DW orientation $\delta=45^\circ$. Furthermore, these bound states exhibit an alignment-dependent chiral spin splitting, a feature they share with bulk magnons.

\begin{figure}[t]
	\includegraphics[width=1\columnwidth]{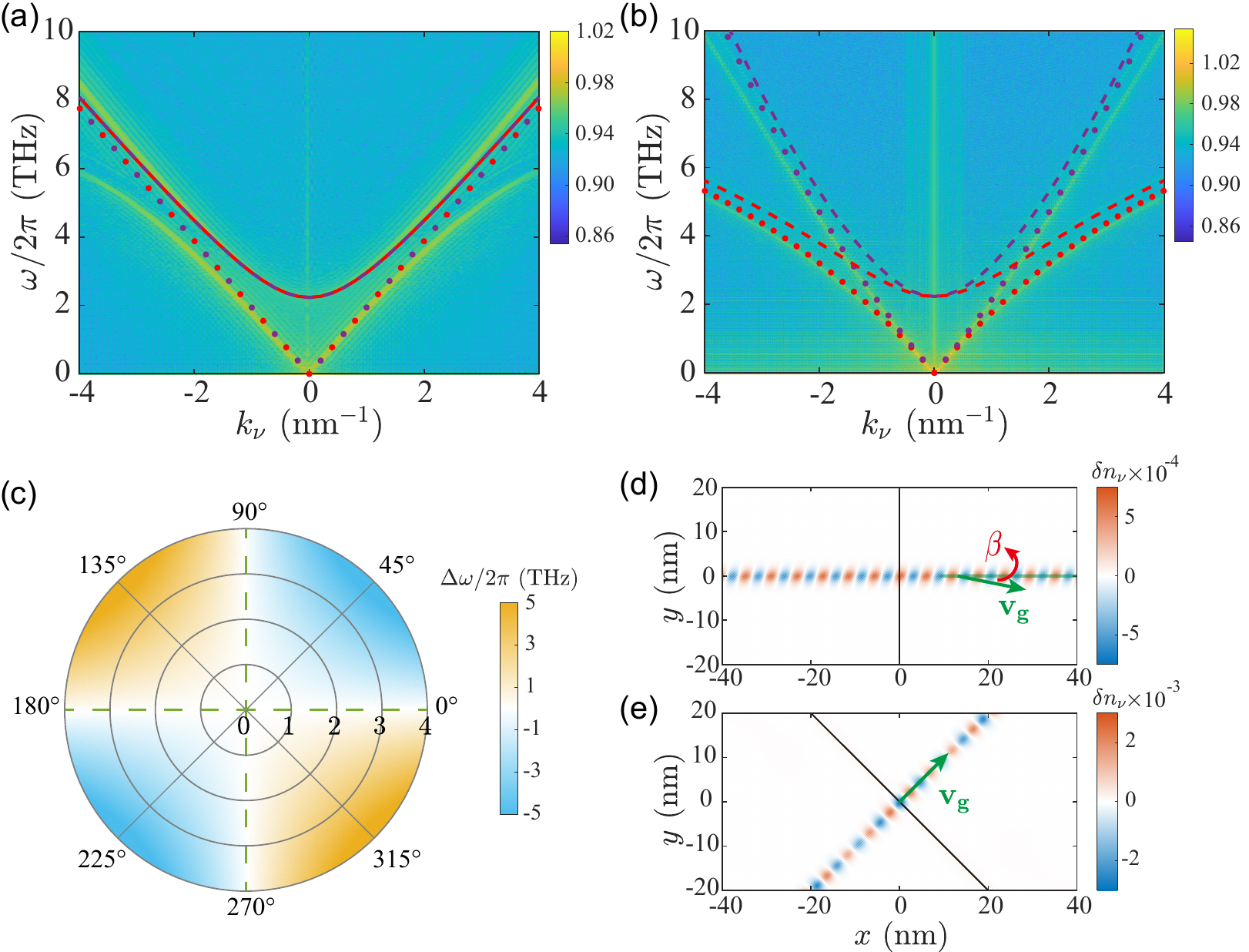}%
	\caption{\label{Fig:2} Magnon dispersion with $\delta = 0^\circ$ (a) and $ 45^\circ$ (b). The color maps are obtained by atomistic spin dynamics simulations. Dots represent the bound magnons from Eq.~\eqref{eq:8} and dashed lines label the bulk magnons from Eq.~\eqref{eq:5}. Red and purple indicate the RH and LH chirality, respectively. (c) The distribution of the frequency splitting $ \Delta \omega/2\pi = (\omega_+-\omega_-)/2\pi $ with respect to the wavenumber $k_\nu$ (unit: nm$^{-1}$) and the oblique angle $\delta$. The dashed green lines indicate the locations where the RH and LH magnons are degenerate. Wavefunctions of bound magnons with $\delta = 0^\circ$ (d) and $45^\circ$ (e) at the $ \omega /2 \pi = 2.0 $ THz. The green vectors represent the group velocity $ {\bf{v}}_g $ with $ \beta $ denoting its misalignment angle with respect to the phase velocity parallel to $ \bm{\nu} $. The black lines represent excitation region.}
\end{figure}

To visualize this behavior, we apply a sinc-function alternating magnetic field to simulate the magnon dispersion, as shown by the color maps in Figs.~\ref{Fig:2}(a) and \ref{Fig:2}(b). It is evident that magnons with different chiralities propagating along the $ x $-axis ($ \delta=0^{\circ} $) are degenerate, whereas they are split with $\delta = 45^\circ$. This feature applies to both bulk and bound magnons. The dots and dashed lines in Figs.~\ref{Fig:2}(a) and \ref{Fig:2}(b) represent the analytical results [Eqs. \eqref{eq:5} and \eqref{eq:8}], which agree excellently with the full micromagnetic simulations.

\begin{figure}[t]
	\includegraphics[width=1\columnwidth]{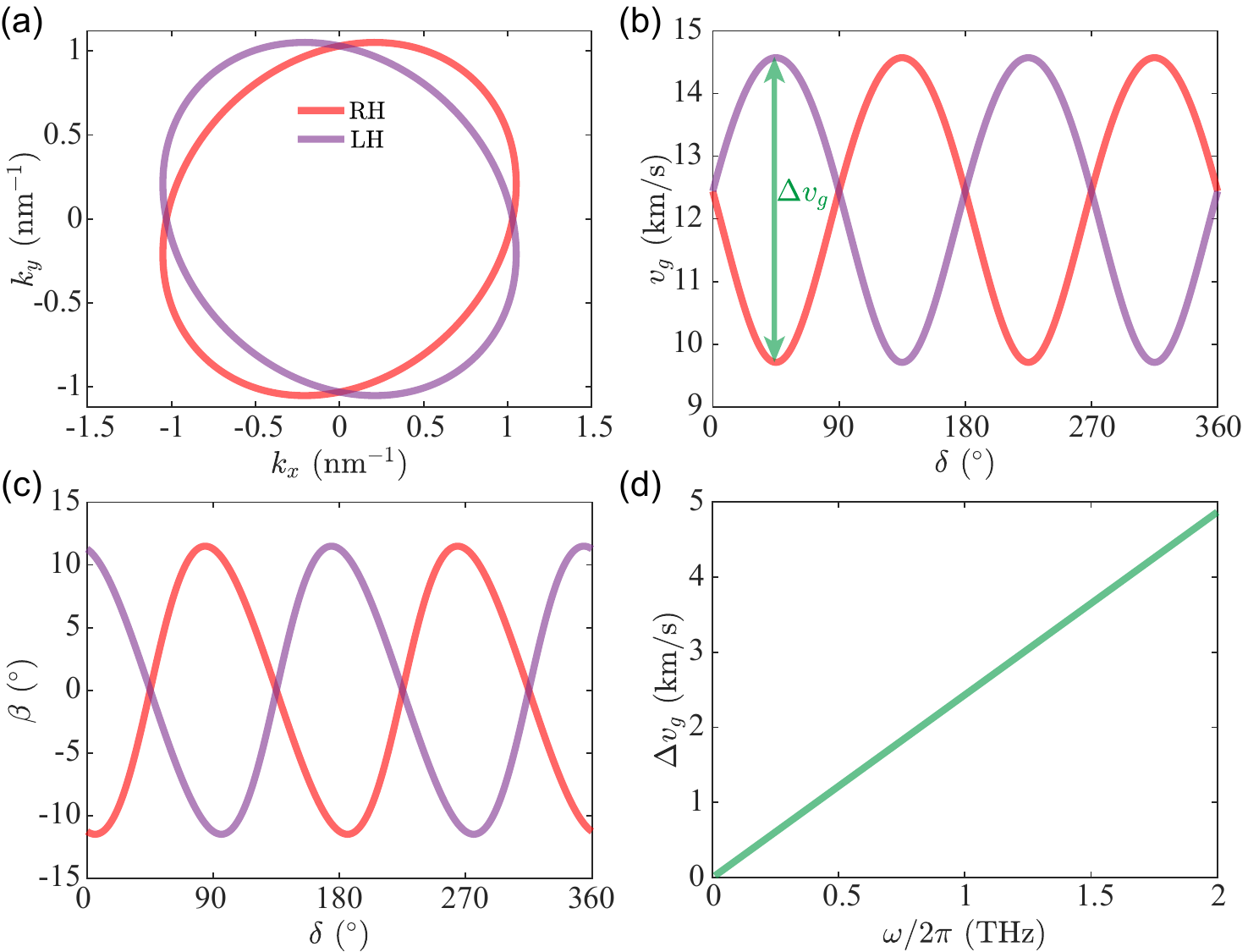}%
	\caption{\label{Fig:3} (a) The isofrequency curves of bound magnons at the $ \omega /2 \pi = 2.0 $ THz. The magnitude $ v_g $ (b) and tilted angle $ \beta $ (c) of the magnon group velocity as a function of DW orientation $ \delta $. Red and purple curves indicate the RH and LH magnons, respectively. (d) The oscillating amplitude of the magnon group velocity $ \Delta v_g $ as a function of its frequency $ \omega/2\pi $ for $\delta = 45^\circ$.}
\end{figure}

Next, we focus on the chiral splitting of bound magnons with respect to both the wavenumber $k_\nu$ and oblique angle $\delta$. It is found that the degenerate points of RH and LH magnons locate on the $ x $ and $ y $ axes with $ \sin\left(2\delta\right)=0$, as plotted by the dashed green line in Fig.~\ref{Fig:2}(c). The spectra of bound magnons in AMDWs demonstrate the $ [C_2||C_{4z}] $ symmetry, which may serve as the most important feature to identify the altermagnets, particularly, in the vicinity of $ k_\nu = 0 $, where $ \omega_\pm/2\pi $ and $ \Delta \omega/2\pi $ reach the gigahertz and megahertz range (see Sec. IV \cite{SM}), respectively. These values are detectable using techniques like all-electrical spin-wave spectroscopy and Brillouin light scattering \cite{yu2021magnetic,chen2025observation}.

To observe the bound magnons directly, we excite RH magnons using a circularly polarized alternating magnetic field ${\bf h}=h_{0}\sin\left(\omega t\right){\bm{\nu}}+h_{0}\cos\left(\omega t\right){\bf e}_{z} $ with $ h_0 = 200 $ mT and $ \omega/2\pi = 2.0 $ THz (below the bulk gap). Figures \ref{Fig:2}(d) and \ref{Fig:2}(e) demonstrate the wavefunction of bound magnons with $\delta$ being $ 0^\circ$ and $45^\circ$, respectively. In both cases, the magnons are confined within the AMDWs. However, for $\delta = 0^\circ$, the wavefunction tilts slightly due to a misalignment between the phase velocity (along the $x$-direction) and the group velocity $ {\bf{v}}_g = \gamma/M_{s} \left(\sqrt{A_{0}A_{1}},A_{2}k_{\nu}\right) $. The tilting angle is $ \beta\approx-11.27^{\circ} $, as illustrated in Fig.~\ref{Fig:2}(d) with green vectors indicating $\mathbf{v}_g$ and black lines marking the excitation region. This tilting stems from the anisotropic exchange term $A_2\sin\left(2\delta\right)$ in Eq.~\eqref{eq:8}, which breaks rotational symmetry, and thus results in a misalignment between the phase and group velocity.
\begin{figure}[htbp]
	\includegraphics[width=1\columnwidth]{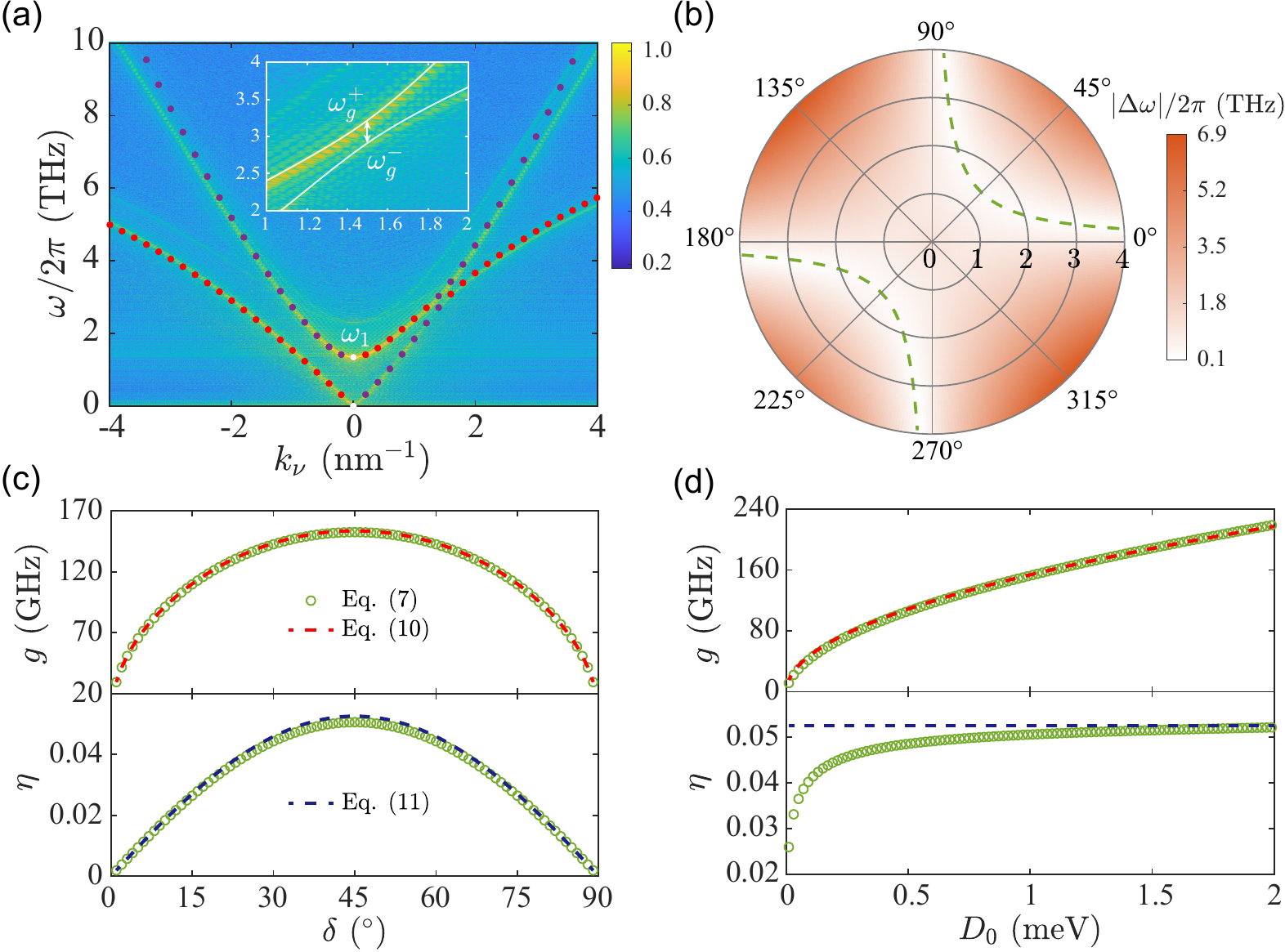}%
	\caption{\label{Fig:4} (a) Magnon dispersion with $\delta = 45^\circ$. The dots represent bound magnons from Eq.~\eqref{eq:7}. Red, purple, and white dots indicate the RH, LH, and linearly polarized magnons, respectively. Inset: zoom in the level anticrossing. (b) The distribution of the frequency splitting $ |\Delta\omega|/2\pi $. The dashed green lines indicate the locations where the coupling strength $ g $ is determined. (c) The coupling strength $g$ and efficiency $\eta$ as a function of the oblique angle $\delta$. In the above calculations, we set $D_0=1$ meV. (d) $ g $ and $\eta$ vs. $D_0$ with $\delta = 45^\circ$. Dots represent numerical results from Eq.~\eqref{eq:7}. Red and blue dotted lines correspond to analytical results from Eqs.~\eqref{eq:10} and \eqref{eq:11}, respectively.}
\end{figure}

The isofrequency curves of bound magnons at $\omega / 2\pi = 2.0$ THz, shown in Fig.~\ref{Fig:3}(a), exhibit strong anisotropy and chiral splitting, consistent with bulk modes \cite{gomonay2024structure}. Here, $ k_{x} = k_\nu \cos\delta $ and $ k_y = k_\nu \sin\delta $. The group velocity magnitude $v_g$ oscillates with $\delta$ (period $\pi$), as plotted in Fig.~\ref{Fig:3}(b). When $ \delta = 0^\circ $, $v_g = 12.44$ km/s for both RH and LH magnons (degenerate). When $ \delta = 45^\circ $, $v_g = 9.71$ km/s (LH) and $14.57$ km/s (RH), showing splitting. Moreover, it can be seen that $ \beta = \arctan\left(v_g^y/v_g^x\right)-\delta $ reaches $ \pm 11.27^\circ $ with $ \delta = 0^\circ $, and $ 0^{\circ} $ for $ \delta = 45^\circ $, as shown in Fig.~\ref{Fig:3}(c). The oscillation amplitude of the group velocity, defined as $ \Delta v_g = \left|v_{g+}\left(\delta=45^\circ\right)-v_{g-}\left(\delta=45^\circ\right)\right|$, increases linearly with frequency $ \omega/2\pi $, as plotted in Fig.~\ref{Fig:3}(d).

\textit{Role of DMI---}The DMI significantly influences the behavior of bound magnons in AMDWs. To see this, we define two new variables $ \phi^{\pm}=n_{1} \mp i n_{2} $, and re-write Eq.~\eqref{eq:6} as
\begin{equation}
	\left(\begin{array}{cc}
	\Omega_{1}-\Omega_{2}-G & G\\
	G & \Omega_{1}+\Omega_{2}-G
	\end{array}\right)\left(\begin{array}{c}
	\phi^{+}\\
	\phi^{-}
	\end{array}\right)=0,
	\label{eq:9}
\end{equation}
where $ \Omega_{1} = -T\omega^{2}+B_{1}k_{\nu}^{2} $, $ \Omega_{2} = -B_{2}\omega k_{\nu}^{2}+D_{1}k_{\nu} $, and $ G = D_{2}/2 $. One can immediately note that the off-diagonal term $ G = \pi\gamma D/\left(8M_{s}w\right) $ leads to the magnon-magnon coupling. The color map in Fig.~\ref{Fig:4}(a) illustrates the dispersion of magnons in AMDWs with $ D_0 = 1 $ meV. A distinctive feature is the unidirectional anticrossing at $ k_\nu >0 $, induced by DMI, which couples RH and LH modes asymmetrically due to the chiral wall texture and hybridizes the bound magnon polarizations (see Fig. S3 \cite{SM}).

The angle-dependence of the chiral splitting $ |\Delta\omega|/2\pi $ is depicted in Fig.~\ref{Fig:4}(b). The strength of magnon-magnon coupling is defined as $g = \min \Big\{ |\Delta\omega|/4\pi \Big\}=\left(\omega_g^+ - \omega_g^-\right)/4\pi$ where the frequency difference reaches its minimum at the wavenumber $k_{\nu 0}$ \cite{liensberger2019exchange}. The dashed green lines in Fig.~\ref{Fig:4}(b) label the contour where $g$ is determined.

Using perturbation theory, we derive the analytical formula of the magnon-magnon coupling \cite{SM}
\begin{equation}
	g\approx\frac{1}{8}\left(\frac{\gamma^{2}A_{0}D}{\pi w M_{s}^{2}}\right)^{1/2}/\left[1-\frac{4w\sqrt{A_{0}A_{1}}}{A_{2}\sin\left(2\delta\right)}\right]^{1/2}.
	\label{eq:10}
\end{equation}
The coupling efficiency is quantified by the following parameter (see Appendix A)
\begin{equation}
	\eta\approx1/\left[1-\frac{4w\sqrt{A_{0}A_{1}}}{A_{2}\sin\left(2\delta\right)}\right].
	\label{eq:11}
\end{equation}
Analytical formula (dashed curves) compares well with numerical results (circles), as shown in Fig.~\ref{Fig:4}(c). The top panel of Fig.~\ref{Fig:4}(d) confirms the square-root dependence of $g$ on $D_0$. However, our formula Eq.~\eqref{eq:11} can only capture of the asymptotic behavior of $\eta$ on $D_0$, as seen in the bottom panel in Fig.~\ref{Fig:4}(d). This is due to the oversimplified treatment in our perturbation calculations.

It is worth pointing out that the location of level anticrossing is limited by the energy gap $ \omega_1=\sqrt{\pi\gamma^{2}A_{0}D/\left(4w M_{s}^{2}\right)} $ of the high-frequency magnon band shown in Fig.~\ref{Fig:4}(a). To further shift the splitting to lower frequencies/wavenumbers, one could use materials with weaker homogeneous exchange (smaller $ |A_0| $), by tuning $ J_1 $ via strain or doping in candidates like $ \rm{Cr_2Te_2O} $, or increase DW width $ w $ (via smaller $ K $ or larger $ A_1 $), or use smaller DMI. These could reduce splitting onset to sub-GHz, probeable by broadband ferromagnetic resonance or microwave spectroscopy.

A key challenge is then how to effectively control the DW orientation. To address this, we employ the SOT to tune the alignment of AMDWs in an (insulating) altermagnetic ring structure \cite{klaui2003domain,sohn2015electrically,hu2016fast}. Our simulations show that the SOT efficiently rotates the wall from $ \delta=0^\circ $ to $ 60^\circ $ within $ 1 $ ns (see Fig. S5 \cite{SM}). Meanwhile, the STT can also be utilized to tune the alignment of AMDWs in a metallic altermagnet ($ \rm{e.g., RuO_2} $), as shown in Fig. S6 \cite{SM}.

\textit{Discussion---}Our work highlights strong coupling between bounded magnon states, an area that remains underexplored \cite{shiota2020tunable}. Although our analysis centers on $d$-wave altermagnets, we expect these findings to extend to other altermagnetic classes, such as $g$- and $i$-wave altermagnets. We numerically find that magnons can propagate along curved AMDWs without scattering, as evidenced in Fig. S7 \cite{SM}. This scatter-free propagation could inspire the development of innovative magnonic circuits \cite{flebus20242024} that exploit the unique properties of altermagnetic textures. Looking ahead, the influence of dipolar interactions and magnon-photon coupling warrants further investigation, as these effects are likely to play significant roles in practical applications.

To verify our theoretical predictions, the following challenges need to be considered: Altermagnets like $ \rm{Cr_2Te_2O} $ should be interfaced with heavy metals (e.g., Pt) for strong SOT, but lattice mismatch could introduce defects affecting wall stability; SOT requires current density as high as $ 10^{11} $ A/m$ ^{2} $, risking Joule heating in nanoscale devices, mitigated by pulsed currents or efficient materials; Defects or roughness may hinder smooth rotation, requiring ultra-clean interfaces. van der Waals heterostructures would serve for better integration; Real-time monitoring of wall orientation via magneto-optical Kerr effect or electrical signals (e.g., anomalous Hall) is needed but challenging at nanoscale.

\textit{Conclusion---}In summary, we have explored the properties of magnons confined within AMDWs, revealing their unique alignment-dependent transport and interactions. We found that the bound magnons exhibit gapless chiral spin splitting akin to bulk magnons in the absence of the DMI. However, their dispersion is distinctly sensitive to the domain wall's orientation relative to the crystal. The DMI induces unidirectional strong coupling between bound magnons. This results in hybrid magnon modes with controllable chiral properties. We finally demonstrated that the orientation of AMDW can be efficiently controlled by electrical means, like SOT and/or STT, enabling on-demand manipulation of the magnon properties. Our findings enhance the understanding of chiral magnon dynamics confined in magnetic textures, highlighting the role of symmetry and interfacial interactions in altermagnets. The ability to tune magnon properties via domain wall orientation and electrical currents paves the way for innovative magnonic devices, leveraging the distinct features of altermagnetism.
\begin{acknowledgments}
 We thank J. Liu, Y. Liu, Y. Su, and Y. Cao for helpful discussions. This work was funded by the National Key R$\&$D Program under Contract No. 2022YFA1402802, the National Natural Science Foundation of China (NSFC) (Grants No. 12374103 and No. 12434003), and Sichuan Science and Technology Program (No. 2025NSFJQ0045). Z. Z. acknowledges the financial support from NSFC (Grant No. 125B2070). Z. J. acknowledges the financial support from NSFC (Grant No. 12404125) and the China Postdoctoral Science Foundation (Grant No. 2024M750337).
\end{acknowledgments}

\begin{thebibliography}{99}
	\newcommand{\DOI}[1]{doi: \href{https://doi.org/#1}{#1}}
	
	\bibitem{vsmejkal2022beyond}L. \v{S}mejkal, J. Sinova, and T. Jungwirth, Beyond conventional ferromagnetism and antiferromagnetism: A phase with nonrelativistic spin and crystal rotation symmetry, \href{https://doi.org/10.1103/PhysRevX.12.031042}{Phys. Rev. X \textbf{12}, 031042 (2022).}
	
	\bibitem{vsmejkal2022emerging}L. \v{S}mejkal, J. Sinova, and T. Jungwirth, Emerging research landscape of altermagnetism, \href{https://doi.org/10.1103/PhysRevX.12.040501}{Phys. Rev. X \textbf{12}, 040501 (2022).}
	
	\bibitem{mazin2022altermagnetism}I. Mazin, Altermagnetism-a new punch line of fundamental magnetism, \href{https://doi.org/10.1103/PhysRevX.12.040002}{Phys. Rev. X \textbf{12}, 040002 (2022).}
	
	\bibitem{bai2024altermagnetism}L. Bai, W. Feng, S. Liu \emph{et al}., Altermagnetism: exploring new frontiers in magnetism and spintronics, \href{https://doi.org/10.1002/adfm.202409327}{Adv. Funct. Mater. \textbf{34}, 2409327 (2024).}
	
	\bibitem{song2025altermagnets}C. Song, H. Bai, Z. Zhou \emph{et al}., Altermagnets as a new class of functional materials, \href{https://doi.org/10.1038/s41578-025-00779-1}{Nat. Rev. Mater. \textbf{10}, 473 (2025).}
	
	\bibitem{vsmejkal2020crystal}L. \v{S}mejkal, R. Gonz{\'a}lez-Hern{\'a}ndez, T. Jungwirth \emph{et al}., Crystal time-reversal symmetry breaking and spontaneous Hall effect in collinear antiferromagnets, \href{https://doi.org/10.1126/sciadv.aaz8809}{Sci. Adv. \textbf{6}, eaaz8809 (2020).}
	
	\bibitem{gonzalez2023spontaneous}R. D. Gonzalez Betancourt, J. Zub{\'a}{\v{c}}, R. Gonzalez-Hernandez \emph{et al}., Spontaneous anomalous Hall effect arising from an unconventional compensated magnetic phase in a semiconductor, \href{https://doi.org/10.1103/PhysRevLett.130.036702}{Phys. Rev. Lett. \textbf{130}, 036702 (2023).}

	\bibitem{reichlova2024observation}H. Reichlova, R. L. Seeger, R. Gonz{\'a}lez-Hern{\'a}ndez \emph{et al}., Observation of a spontaneous anomalous Hall response in the $\mathrm{Mn_5Si_3}$ $d$-wave altermagnet candidate, \href{https://doi.org/10.1038/s41467-024-48493-w}{Nat. Commun. \textbf{15}, 4961 (2024).}
	
	\bibitem{badura2024observation}A. Badura, W. H. Campos, V. K. Bharadwaj \emph{et al}., Observation of the anomalous Nernst effect in altermagnetic candidate $\mathrm{Mn_5Si_3}$, \href{https://doi.org/10.48550/arXiv.2403.12929}{arXiv:2403.12929 (2024).}

	\bibitem{han2024observation}L. Han, X. Fu, W. He \emph{et al}., Observation of non-volatile anomalous Nernst effect in altermagnet with collinear N\'eel vector, \href{https://doi.org/10.48550/arXiv.2403.13427}{arXiv:2403.13427 (2024).}
	
	\bibitem{vsmejkal2022giant}L. {\v{S}}mejkal, A. B. Hellenes, R. Gonz{\'a}lez-Hern{\'a}ndez \emph{et al}., Giant and tunneling magnetoresistance in unconventional collinear antiferromagnets with nonrelativistic spin-momentum coupling, \href{https://doi.org/10.1103/PhysRevX.12.011028}{Phys. Rev. X \textbf{12}, 011028 (2022).}
	
	\bibitem{liu2024giant}F. Liu, Z. Zhang, X. Yuan \emph{et al}., Giant tunneling magnetoresistance in insulated altermagnet/ferromagnet junctions induced by spin-dependent tunneling effect, \href{https://doi.org/10.1103/PhysRevB.110.134437}{Phys. Rev. B \textbf{110}, 134437 (2024).}
	
	\bibitem{naka2019spin}M. Naka, S. Hayami, H. Kusunose \emph{et al}., Spin current generation in organic antiferromagnets, \href{ https://doi.org/10.1038/s41467-019-12229-y}{Nat. Commun. \textbf{10}, 4305 (2019).}
	
	\bibitem{bai2022observation}H. Bai, L. Han, X. Y. Feng \emph{et al}., Observation of Spin Splitting Torque in a Collinear Antiferromagnet $\mathrm{RuO_2}$, \href{https://doi.org/10.1103/PhysRevLett.128.197202}{Phys. Rev. Lett. \textbf{128}, 197202 (2022).}
	
	\bibitem{karube2022observation}S. Karube, T. Tanaka, D. Sugawara \emph{et al}., Observation of Spin-Splitter Torque in Collinear Antiferromagnetic $\mathrm{RuO_2}$, \href{https://doi.org/10.1103/PhysRevLett.129.137201}{Phys. Rev. Lett. \textbf{129}, 137201 (2022).}
	
	\bibitem{baltz2018antiferromagnetic}V. Baltz, A. Manchon, M. Tsoi \emph{et al}., Antiferromagnetic spintronics, \href{https://doi.org/10.1103/RevModPhys.90.015005}{Rev. Mod. Phys. \textbf{90}, 015005 (2018).}
	
	\bibitem{vsmejkal2023chiral}L. {\v{S}}mejkal, A. Marmodoro, K. H. Ahn \emph{et al}., Chiral Magnons in Altermagnetic $\mathrm{RuO_2}$, \href{https://doi.org/10.1103/PhysRevLett.131.256703}{Phys. Rev. Lett. \textbf{131}, 256703 (2023).}
		
	\bibitem{cui2023efficient}Q. Cui, B. Zeng, P. Cui \emph{et al}., Efficient spin Seebeck and spin Nernst effects of magnons in altermagnets, \href{https://doi.org/10.1103/PhysRevB.108.L180401}{Phys. Rev. B \textbf{108}, L180401 (2023).}
	
	\bibitem{liu2024chiral}Z. Liu, M. Ozeki, S. Asai \emph{et al}., Chiral split magnon in altermagnetic MnTe, \href{https://doi.org/10.1103/PhysRevLett.133.156702}{Phys. Rev. Lett. \textbf{133}, 156702 (2024).}
	
	\bibitem{morano2025absence}V. C. Morano, Z. Maesen, S. E. Nikitin \emph{et al}., Absence of Altermagnetic Magnon Band Splitting in $\mathrm{MnF_2}$, \href{https://doi.org/10.1103/PhysRevLett.134.226702}{Phys. Rev. Lett. \textbf{134}, 226702 (2025).}
	
	\bibitem{gomonay2024structure}O. Gomonay, V. P. Kravchuk, R. Jaeschke-Ubiergo \emph{et al}., Structure, control, and dynamics of altermagnetic textures, \href{https://doi.org/10.1038/s44306-024-00042-3}{npj Spintronics \textbf{2}, 35 (2024).}
	
	\bibitem{krempasky2024altermagnetic}J. Krempask{\`y}, L. {\v{S}}mejkal, S. W. D'souza \emph{et al}., Altermagnetic lifting of Kramers spin degeneracy, \href{https://doi.org/10.1038/s41586-023-06907-7}{Nature \textbf{626}, 517 (2024).}
	
	\bibitem{amin2024nanoscale}O. J. Amin, A. Dal Din, E. Golias \emph{et al}., Nanoscale imaging and control of altermagnetism in MnTe, \href{https://doi.org/10.1038/s41586-024-08234-x}{Nature \textbf{636}, 348 (2024).}
	
	\bibitem{ding2024large}J. Ding, Z. Jiang, X. Chen \emph{et al}., Large band-splitting in $g$-wave type altermagnet CrSb, \href{https://doi.org/10.1103/PhysRevLett.133.206401}{Phys. Rev. Lett. \textbf{133}, 206401 (2024).}
	
	\bibitem{hariki2024xray}A. Hariki, A. Dal Din, O. J. Amin \emph{et al}., X-Ray Magnetic Circular Dichroism in Altermagnetic $\alpha$-MnTe, \href{https://doi.org/10.1103/PhysRevLett.132.176701}{Phys. Rev. Lett. \textbf{132}, 176701 (2024).}
	
	\bibitem{hiraishi2024nonmagnetic}M. Hiraishi, H. Okabe, A. Koda \emph{et al}., Nonmagnetic ground state in $\mathrm{RuO_2}$ revealed by muon spin rotation, \href{https://doi.org/10.1103/PhysRevLett.132.166702}{Phys. Rev. Lett. \textbf{132}, 166702 (2024).}
	
	\bibitem{liu2024absence}J. Liu, J. Zhan, T. Li \emph{et al}., Absence of Altermagnetic Spin Splitting Character in Rutile Oxide $\mathrm{RuO_2}$, \href{https://doi.org/10.1103/PhysRevLett.133.176401}{Phys. Rev. Lett. \textbf{133}, 176401 (2024).}
	
	\bibitem{noh2025tunneling}S. Noh, G. H. Kim, J. Lee \emph{et al}., Tunneling Magnetoresistance in Altermagnetic $\mathrm{RuO_2}$-Based Magnetic Tunnel Junctions, \href{https://doi.org/10.1103/nrk5-5zrj}{Phys. Rev. Lett. \textbf{134}, 246703 (2025).}
		
	\bibitem{yu2021magnetic}H. Yu, J. Xiao, and H. Schultheiss, Magnetic texture based magnonics, \href{https://doi.org/10.1016/j.physrep.2020.12.004}{Phys. Rep. \textbf{905}, 1 (2021).}
	
	\bibitem{jin2025strong}Z. Jin, T. Gong, J. Liu \emph{et al}., Strong coupling of chiral magnons in altermagnets, \href{https://doi.org/10.1103/gn6c-1q19}{Phys. Rev. Lett. \textbf{135}, 126702 (2025).}	

	\bibitem{Goldstone2022}M. Grassi, M. Geilen, K. A. Oukaci \emph{et al}., Higgs and Goldstone spin-wave modes in striped magnetic texture, \href{https://doi.org/10.1103/PhysRevB.105.094444}{Phys. Rev. B \textbf{105}, 094444 (2022).}
			
	\bibitem{kolezhuk2024current}O. Kolezhuk, R. Teslia, I. Buryak \emph{et al}., Current-controlled chirality dynamics in a mesoscopic magnetic domain wall, \href{https://doi.org/10.1103/PhysRevB.109.134418}{Phys. Rev. B \textbf{109}, 134418 (2024).}
	
	\bibitem{zarzuela2025transport}R. Zarzuela, R. Jaeschke-Ubiergo, O. Gomonay \emph{et al}., Transport theory and spin-transfer physics in $d$-wave altermagnets, \href{https://doi.org/10.1103/PhysRevB.111.064422}{Phys. Rev. B \textbf{111}, 064422 (2025).}
	
	\bibitem{kravchuk2025chiral}V. P. Kravchuk, K. V. Yershov, J. I. Facio \emph{et al}., Chiral magnetic excitations and domain textures of $g$-wave altermagnets, \href{https://doi.org/10.48550/arXiv.2504.05241}{arXiv:2504.05241 (2025).}
	
	\bibitem{garcia2015narrow}F. Garcia-Sanchez, P. Borys, R. Soucaille \emph{et al}., Narrow magnonic waveguides based on domain walls, \href{https://doi.org/10.1103/PhysRevLett.114.247206}{Phys. Rev. Lett. \textbf{114}, 247206 (2015).}
	
	\bibitem{zhang2018eavesdropping}B. Zhang, Z. Wang, Y. Cao \emph{et al}., Eavesdropping on spin waves inside the domain-wall nanochannel via three-magnon processes, \href{https://doi.org/10.1103/PhysRevB.97.094421}{Phys. Rev. B \textbf{97}, 094421 (2018).}
	
	\bibitem{yu2018polarization}W. Yu, J. Lan, and J. Xiao, Polarization-selective spin wave driven domain-wall motion in antiferromagnets, \href{https://doi.org/10.1103/PhysRevB.98.144422}{Phys. Rev. B \textbf{98}, 144422 (2018).}
	
	\bibitem{henry2019unidirectional}Y. Henry, D. Stoeffler, J. V. Kim \emph{et al}., Unidirectional spin-wave channeling along magnetic domain walls of Bloch type, \href{https://doi.org/10.1103/PhysRevB.100.024416}{Phys. Rev. B \textbf{100}, 024416 (2019).}
	
	\bibitem{park2019channeling}H. K. Park and S. K. Kim, Channeling of spin waves in antiferromagnetic domain walls, \href{https://doi.org/10.1103/PhysRevB.103.214420}{Phys. Rev. B \textbf{103}, 214420 (2019).}
			
	\bibitem{wang2020stacking}C. Wang, Y. Gao, H. Lv \emph{et al}., Stacking domain wall magnons in twisted van der Waals magnets, \href{https://doi.org/10.1103/PhysRevLett.125.247201}{Phys. Rev. Lett. \textbf{125}, 247201 (2020).}
	
	\bibitem{chen2020narrow}G. Chen, J. Lan, T. Min \emph{et al}., Narrow Waveguide Based on Ferroelectric Domain Wall, \href{https://doi.org/10.1088/0256-307X/38/8/087701}{Chin. Phys. Lett. \textbf{38}, 087701 (2020).}
	
	\bibitem{qiu2022tunable}L. Qiu and K. Shen, Tunable spin-wave nonreciprocity in synthetic antiferromagnetic domain walls, \href{https://doi.org/10.1103/PhysRevB.105.094436}{Phys. Rev. B \textbf{105}, 094436 (2022).}
	
	\bibitem{liang2022nonreciprocal}X. Liang, Z. Wang, P. Yan \emph{et al}., Nonreciprocal spin waves in ferrimagnetic domain-wall channels, \href{https://doi.org/10.1103/PhysRevB.106.224413}{Phys. Rev. B \textbf{106}, 224413 (2022).}
	
	\bibitem{li2024realizing}Z. Li, X. Liu, Z. Yan \emph{et al}., Realizing polarization-dependent unidirectional magnon
	channel in antiferromagnetic domain wall, \href{https://doi.org/10.1063/5.0181317}{Appl. Phys. Lett. \textbf{124}, 032401 (2024).}
				
	\bibitem{wagner2016magnetic}K. Wagner, A. K{\'a}kay, K. Schultheiss \emph{et al}., Magnetic domain walls as reconfigurable spin-wave nanochannels, \href{https://doi.org/10.1038/nnano.2015.339}{Nat. Nanotechnol. \textbf{11}, 432 (2016).}
	
	\bibitem{albisetti2018nanoscale}E. Albisetti, D. Petti, G. Sala \emph{et al}., Nanoscale spin-wave circuits based on engineered reconfigurable spin-textures, \href{https://doi.org/10.1038/s42005-018-0056-x}{Commun Phys \textbf{1}, 56 (2018).}
	
	\bibitem{sluka2019emission}V. Sluka, T. Schneider, R. A. Gallardo \emph{et al}., Emission and propagation of 1D and 2D spin waves with nanoscale wavelengths in anisotropic spin textures, \href{https://doi.org/10.1038/s41565-019-0383-4}{Nat. Nanotechnol. \textbf{14}, 328-333 (2019).}
	
	\bibitem{chen2025observation}J. Chen, Z. Jin, R. Yuan \emph{et al}., Observation of Coherent Gapless Magnons in an Antiferromagnet, \href{https://doi.org/10.1103/PhysRevLett.134.056701}{Phys. Rev. Lett. \textbf{134}, 056701 (2025).}

	\bibitem{SM}See Supplemental Material at http://link.aps.org/supplemental/ for the continuum approximation in altermagnets (Sec. I), the dispersions of bound magnons in AMDWs (Sec. II), the details of atomistic spin dynamic simulations with AMDWs (Sec. III), the dispersion of magnon bound states in the vicinity of $ k_\nu = 0 $ (Sec. IV), the polarization of bound magnons (Sec. V), the magnon-magnon coupling induced by the DMI (Sec. VI), the current induced re-orientation of AMDWs (Sec. VII), and the magnon propagation in curved walls (Sec. VIII), which includes Refs. \cite{baltz2018antiferromagnetic,gomonay2024structure,kim2019tunable,jin2023nonlinear,vakili2025spin}
			
	\bibitem{kim2019tunable}S. K. Kim, K. Nakata, D. Loss \emph{et al}., Tunable magnonic thermal Hall effect in skyrmion crystal phases of ferrimagnets, \href{https://doi.org/10.1103/PhysRevLett.122.057204}{Phys. Rev. Lett. \textbf{122}, 057204 (2019).}
	
	\bibitem{jin2023nonlinear}Z. Jin, X. Yao, Z. Wang \emph{et al}., Nonlinear topological magnon spin Hall effect, \href{https://doi.org/10.1103/PhysRevLett.131.166704}{Phys. Rev. Lett. \textbf{131}, 166704 (2023).}
		
	\bibitem{vakili2025spin}H. Vakili, E. Schwartz, and A. A. Kovalev, Spin-Transfer Torque in Altermagnets with Magnetic Textures, \href{https://doi.org/10.1103/PhysRevLett.134.176401}{Phys. Rev. Lett. \textbf{134}, 176401 (2025).}
	
	\bibitem{fert2013skyrmions}A. Fert, V. Cros, and J. Sampaio, Skyrmions on the track, \href{https://doi.org/10.1038/s41565-019-0383-4}{Nat. Nanotechnol. \textbf{8}, 152 (2013).}
	
	\bibitem{jin2024skyrmion}Z. Jin, Z. Zeng, Y. Cao \emph{et al}., Skyrmion Hall Effect in Altermagnets, \href{https://doi.org/10.1103/PhysRevLett.133.196701}{Phys. Rev. Lett. \textbf{133}, 196701 (2024).}
	
	\bibitem{jiao2024universal}X. Jiao, X. S. Wang, and J. Lan, Universal spin wave driven domain wall velocity in biaxial ferromagnets, \href{https://doi.org/10.1103/PhysRevB.109.094428}{Phys. Rev. B \textbf{109}, 094428 (2024).}
	
	\bibitem{liensberger2019exchange}L. Liensberger, A. Kamra, H. Maier-Flaig \emph{et al}., Exchange-Enhanced Ultrastrong Magnon-Magnon Coupling in a Compensated Ferrimagnet, \href{https://doi.org/10.1103/PhysRevLett.123.117204}{Phys. Rev. Lett. \textbf{123}, 117204 (2019).}
	
	\bibitem{klaui2003domain}M. Kl{\"a}ui, C. A. F. Vaz, J. A. C. Bland\emph{et al}., Domain wall motion induced by spin polarized currents in ferromagnetic ring structures, \href{https://doi.org/10.1063/1.1588736}{Appl. Phys. Lett. \textbf{83}, 105 (2003).}
	
	\bibitem{sohn2015electrically}H. Sohn, M. E. Nowakowski, C. Liang \emph{et al}., Electrically Driven Magnetic Domain Wall Rotation in Multiferroic Heterostructures to Manipulate Suspended On-Chip Magnetic Particles, \href{https://doi.org/10.1021/nn5056332}{ACS Nano \textbf{9}, 4814 (2015).}
	
	\bibitem{hu2016fast}J. M. Hu, T. Yang, K. Momeni \emph{et al}., Fast Magnetic Domain-Wall Motion in a Ring-Shaped Nanowire Driven by a Voltage, \href{https://doi.org/10.1021/acs.nanolett.5b05046}{Nano Lett. \textbf{16}, 2341 (2016).}
	
	\bibitem{shiota2020tunable}Y. Shiota, T. Taniguchi, M. Ishibashi \emph{et al}., Tunable magnon-magnon coupling mediated by dynamic dipolar interaction in synthetic antiferromagnets, \href{https://doi.org/10.1103/PhysRevLett.125.017203}{Phys. Rev. Lett. \textbf{125}, 017203 (2020).}
	
	\bibitem{flebus20242024}B. Flebus, D. Grundler, B. Rana \emph{et al}., The 2024 magnonics roadmap, \href{https://doi.org/10.1088/1361-648X/ad399c}{J. Phys.: Condens. Matter \textbf{36}, 363501 (2024).}

\end{thebibliography}

\end{document}